\documentclass[acmsmall]{acmart}

\usepackage{subcaption}
\usepackage{graphicx}
\usepackage{amsmath,amsfonts}
\usepackage{algorithm}
\usepackage{algpseudocode}
\usepackage{multirow}

\AtBeginDocument{%
  \providecommand\BibTeX{{%
    \normalfont B\kern-0.5em{\scshape i\kern-0.25em b}\kern-0.8em\TeX}}}

\setcopyright{acmcopyright}
\copyrightyear{2018}
\acmYear{2018}
\acmDOI{XXXXXXX.XXXXXXX}

\acmJournal{JACM}
\acmVolume{37}
\acmNumber{4}
\acmArticle{111}
\acmMonth{8}

\begin{document}

\title{Quantitative Analysis of Cultural Dynamics Seen from an Event-based Social Network}

\author{Bayu Adhi Tama}
\affiliation{%
  \institution{University of Maryland, Baltimore County (UMBC)}
  % \streetaddress{1000 Hilltop Circle}
  \city{Baltimore}
  \state{MD}
  \country{USA}}
% \email{bayu@umbc.edu}

\author{Jaehong Kim}
\affiliation{%
  \institution{Korea Advanced Institute of Science \& Technology (KAIST)}
  \city{Daejeon}
  \country{Republic of Korea}
}

\author{Jaehyuk Park}
\affiliation{%
 \institution{KDI School of Public Policy and Management}
 \city{Sejong}
 \country{Republic of Korea}}

\author{Lev Manovich}
\authornote{Corresponding authors}
\affiliation{%
  \institution{City University of New York}
  % \streetaddress{30 Shuangqing Rd}
  \city{NY}
  % \state{Beijing Shi}
  \country{USA}}
  
\author{Meeyoung Cha}
\authornotemark[1]
\affiliation{%
  \institution{Institute for Basic Science}
  % \streetaddress{8600 Datapoint Drive}
  \city{Daejeon}
  % \state{Texas}
  \country{Republic of Korea}
  % \postcode{78229}
  }

% \author{Anonymous}
% \affiliation{%
%   \institution{University of XYZ}
%   % \streetaddress{1000 Hilltop Circle}
%   \city{ABC}
%   \state{DEF}
%   \country{GHI}}
% % \email{bayu@umbc.edu}

\renewcommand{\shortauthors}{Anonymous, et. al.}

\begin{abstract}
Culture is a collection of connected and potentially interactive patterns that characterize a social group or a passed-on idea that people acquire as members of society. While offline activities can provide a better picture of the geographical association of cultural traits than online activities, gathering such data on a large scale has been challenging. Here, we use multi-decade longitudinal records of cultural events from Meetup.com, the largest event-based social networking service, to examine the landscape of offline cultural events. We analyze the temporal and categorical event dynamics driven by cultural diversity using over 2 million event logs collected over 17 years in 90 countries. Our results show that the national economic status explains 44.6 percent of the variance in total event count, while cultural characteristics such as individualism and long-term orientation explain 32.8 percent of the variance in topic categories. Furthermore, our analysis using hierarchical clustering reveals cultural proximity between the topics of socio-cultural activities (e.g., politics, leisure, health, technology). We expect that this work provides a landscape of social and cultural activities across the world, which allows us to better understand their dynamical patterns as well as their associations with cultural characteristics.
\end{abstract}





\maketitle

\section{Introduction}

Social events are spatial-temporal phenomena, where each event has its own unique characteristics of interactions among the participants, topic, description, and geographic region. In particular, a live social event on a certain topic facilitates a unique offline human interaction, as they serve the special purpose of social gathering~\cite{getz2008event}. The expectations, emotions, and attitudes of participants evolve, resulting in a unique human experience, even when events are held on the same topic. Because of such a one-of-a-kind experience, offline live events are appealing, even fascinating, and the organizer's objective is to cultivate a 'once in a lifetime' image for them~\cite{fenich2013meetings,getz2019event}. Furthermore, virtual events, which are conveyed via various social media platforms, provide something of interest and value to people and the tourism industry; they are different types of event experiences~\cite{rojek1997touring,rogers2014diffusion}.

Live events are frequently used to disseminate information. One of the most important characteristics of globalization in the twenty-first century is the adoption of different ideas, interests, beliefs, and lifestyles across cultural boundaries~\cite{BOLI2015225,choudhary2019analysing,frigo2017energy}. Small-scale ethnographic studies can capture local examples of these processes in specific moments. In contrast, computational analysis of social network services, blogs, or websites can provide a global landscape, albeit at the risk of overestimating the impact of globalization~\cite{manovich2020,can2019new,hamidi2011localization}.

This research employs a novel method for studying cultural globalization by using longitudinal data from millions of offline events scheduled through Meetup.com. This platform allows users to communicate virtually with others and meet in person to discuss whatever and wherever they are interested in~\cite{RN23}. The existence of a physical event on a particular topic in a specific location implies that enough people in the area are interested in the topic. The geographic binding of events is perhaps the most critical difference between offline events like those hosted on Meetup and other online communities, which do not restrict who can join and thus give the impression that all ideas are present everywhere. As a result, we assume that offline events, rather than their online counterparts, provide better geo-contextual signals on the topical interest of users in a given region~\cite{manovich2020}.

Meetup.com, as a prominent social media platform specializing in hosting offline cultural events, has its own design and conventions that define how people interact with the platform. Many of its topics, for example, are amusing and attract participants, particularly the civically disengaged. It also allows people to connect regardless of their ethnicity, social class, or geographic location. Book clubs, for example, can unite people of all ages, classes, and places~\cite{RN11,RN10,RN20}. The platform's versatility in terms of how it is used for not only enjoyment but also social involvement and participation is unique. Meetup.com mixes virtual and real elements to enhance social capital from strangers, which makes it one of the best platforms for analyzing cultural events at a regional level.

There has been multidisciplinary research focusing on specific patterns of social media usage, such as the motivations for user participation and consequences~\cite{zheng2016excessive,khan2017social}. Prior research has shown that hashtagging behavior as a form of social media communication can be culturally affected~\cite{sheldon2017cross}. Other works, such as~\cite{sheldon2017cross,hong2018facebook,lin2018cultural}, have discussed the cultural implications of social media communication. Social platforms also enable political participation, ideological categorization, and social movement~\cite{koiranen2020ideological,shen2020examining,yu2018social}. Moreover, the work of~\cite{lee2018does} examines the correlation between social media usage and political polarization through increased political engagement. However, no prior study has investigated the specific link between cross-cultural differences and event-based social networks at the country level due to a lack of empirical data.

\begin{figure}[htpb!]
     \begin{subfigure}[b]{0.8\textwidth}
         \centering
         \includegraphics[width=1\textwidth]{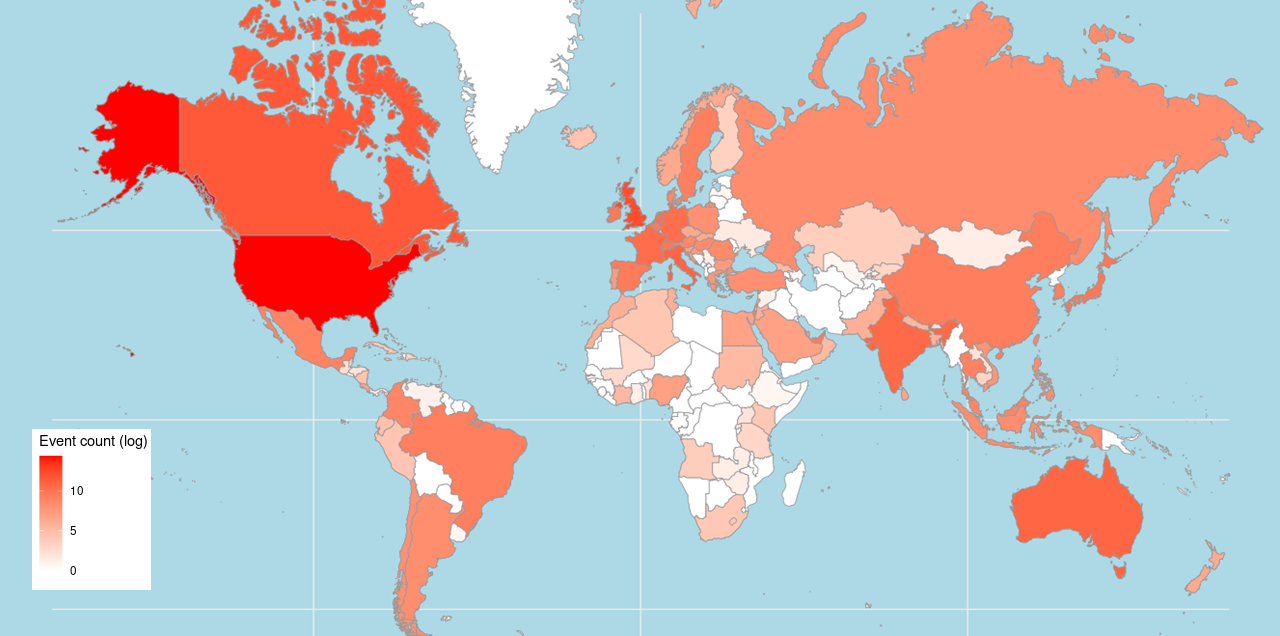}
         \caption{}
         \label{geoplotcities}
     \end{subfigure}
     \begin{subfigure}[b]{0.8\textwidth}
         \centering
         \includegraphics[width=1\textwidth]{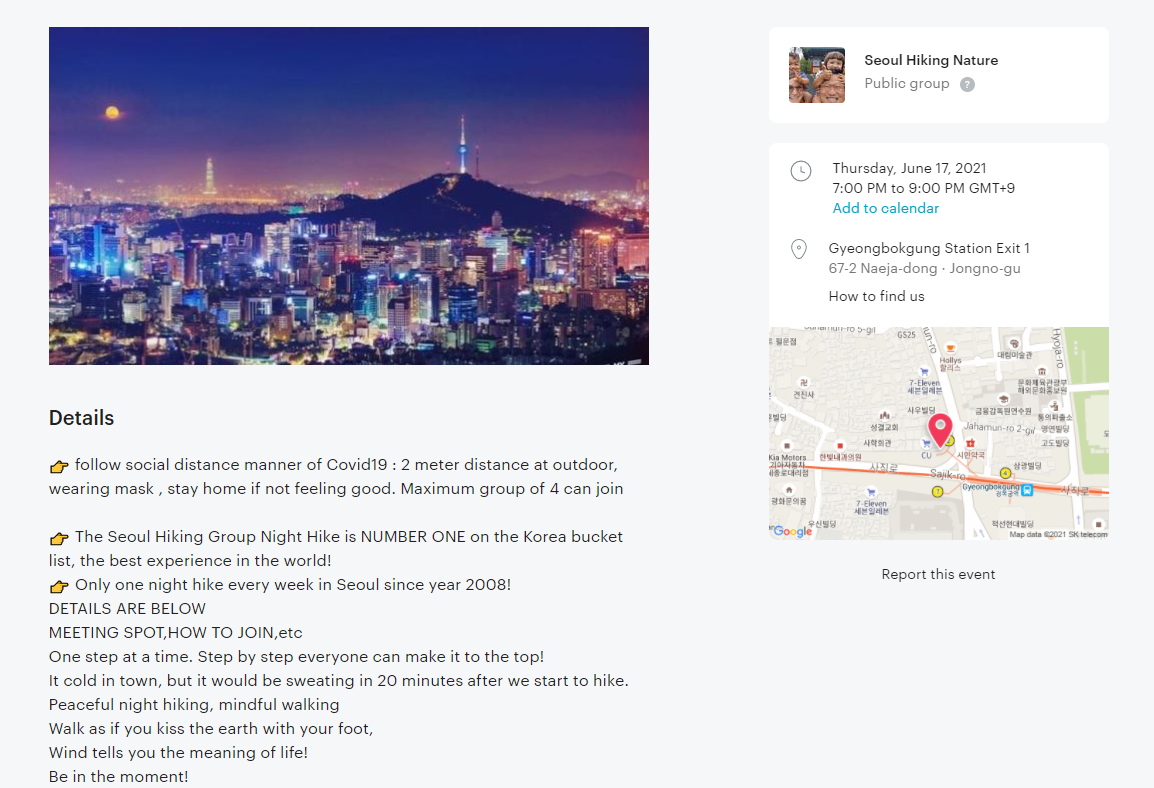}
         \caption{}
         \label{screenshot}
     \end{subfigure}
\caption{(a) Number of events (log-scaled) at country level. Cultural events through Meetup.com are happening across all continents, although more concentrated in North America, Western Europe, and Southeast Asia. (b) A Meetup.com example page for a hiking event in Seoul, which includes information about the event's name, location, date, and description.}
\end{figure}

Here, by proposing an organic approach, we advance the research on the relationship between cultural traits and the event dynamics patterns of Meetup.com in many countries worldwide. Event dynamics of each country are examined using various behavioral features, including the participating group, temporal, and category-related patterns. Furthermore, our approach identifies the hierarchical predominance of event categories across multiple countries and argues that scrutinizing the inter-connected hierarchical structure is a crucial step toward a wider understanding of cultural structures and traits. We show that the category prevalence is shaped by the hierarchical structure of cultural traits, specifically the Hoefstede~\cite{RN66} and Inglehart-Welzel~\cite{inglehart2010wvs} cultural dimension theory. Our dataset of over 2 million events organized in over 17,000 cities in 90 countries span 17 years between 2003 and 2019 (see Figure~\ref{geoplotcities}). The dataset includes its category, date, location, and short text description posted by event organizers, as shown in the main interface (Figure~\ref{screenshot}). Event organizers select the categories from the 33 available categories, which are grouped into eleven mid-level categories and seven top-level categories (see Appendix).

The remainder of the paper is organized as follows. Section~\ref{methods} describes how overall, temporal, and category-related behavior patterns are measured. The results of this study, presented in Section~\ref{find}, include the correlation results and hierarchical structure of the category prevalence. We discuss our findings in Section~\ref{dis}, and lastly, we conclude this paper with some remarks in Section~\ref{conc}.

\section{Methods}
\label{methods}
\subsection{Measuring Overall Behavior Pattern}
Throughout the investigation, an overall behavior pattern is examined via a summary statistic consisting of the total event count (TEC) for the period, the total number of transactions (N$ {Trans}$), the average number of events per transaction ($\mu {Trans}$), and the standard deviation of events per transaction ($\sigma {Trans}$). We specify the transaction table of each country as follows. Let transaction $t$ denote a set of event categories that exist on the same day in a certain nation, $\mathcal{T}$ denote a set of $m$ transactions, $\mathcal{T}=\{t_{1},t_{2},...,t_{m}\}$, denoting a transaction table, and $\mathcal{E}$ denote a set of $n$ different event categories $\mathcal{E}=\{e_{1}, e_{2}, ..., e_{n}\}$. An $\mathcal{S}$ is a collection of event categories, such as $\mathcal{S}\subset \mathcal{E}$. Because the distributions of TEC and N$_{Trans}$ are positively skewed, we apply log transformation to them. We use the coefficient of variation (CoV) to measure the relative variability of a country's event patterns:
\begin{equation}
    CoV=\frac{\sigma_{Trans}}{\mu_{Trans}}
\end{equation}
The country is more likely to have unequal distribution of events across transactions as $CoV$ rises and vice versa.

\subsection{Measuring Temporal Behavior Pattern}
Three time frames, namely, week (W), month (M), and quarter (Q), are used to measure temporal behavior patterns at various granularities. To calculate the variability of a country's event pattern through time windows, we define $\mathcal{E}^{W}=\{e_{Mon},e_{Tue},...,e_{Sun}\}$ as the total of event $e$ produced in each week; $\mathcal{E}^{M}=\{e_{W1},e_{W2},e_{W3},e_{W4}\}$ as the total event $e$ produced in each month; and $\mathcal{E}^{Q}=\{e_{M1},e_{M2},e_{M3}\}$ as the total of event $e$ in each three-month intervals.  We calculate the variability of event patterns in weekly ($\mu_{W}$), monthly ($\mu_{M}$), and quarterly ($\mu_{Q}$)  measurements for each country.

The average cosine similarity coefficients across adjacent time intervals are used to calculate a country's consistency in creating events during a monthly (P$_{M}$) and quarterly (P$_{Q}$) observation period. We calculate the monthly persistence period by aggregating the number of events at the weekly interval (e.g., four elements per month) and obtaining the proportion of each element. Monthly persistence is computed as:
\begin{equation}
    P_{M}=\frac{\sum_{i=0}^{n-1}cos(Z_{i},Z_{i+1})}{n}
\end{equation}
where $Z_{i}$ is the vector representing the relative amount of events created in each monthly interval in a given month $i$, and $n$ is the number of months in a country. Similarly, for the quarterly observation periods,  we construct $P_{Q}$ by aggregating the number of events monthly (i.e., three elements for each quarter). A persistence $P$ value of 1 indicates that the relative number of occurrences remains constant over time periods and vice versa.

Bursty dynamics ($\beta$) is a method of calculating the intensity of a large number of events over a short period of time. The inter-event times are first calculated as the daily difference between two adjacent transactions. Inter-event times are described as $\gamma_{j}=\Gamma_{j}-\Gamma_{j-1}$, where $\Gamma_{j}$ signifies the transaction created at time $j$. The following formula can be used to determine bursty dynamics:
\begin{equation}
    \beta=\frac{\zeta-1}{\zeta+1}
\end{equation}
where $\zeta$ is specified as $\zeta=\frac{\tau}{\sigma}$ with $\tau$ and $\sigma$ indicate the average and standard deviation of the transaction's inter-event timings, respectively. When the burstiness parameter $\beta$ is -1, the country's event pattern is completely steady. The country's event pattern is entirely random when $\beta=0$. Finally, a value of 1 for the parameter $\beta$ results in significant and unexpected increases in the number of occurrences in each transaction.

\subsection{Measuring Category-related Behavior Pattern}
The category-related behavior metrics are calculated using the total number of events to assess the frequency of occurrences in various categories. The number of categories ($N_{Cat}$) indicates how many separate categories were covered during the period. In addition, we assess the diversity of events produced by each country by examining the diversity of categories, $D_{Cat}$, as defined by the Shannon-Wiener diversity index's relative score:
\begin{equation}
    D_{Cat}(c)=-\frac{\sum_{k=1}^{N_{Cat}}p_{kc}log(p_{kc})}{log N_{Cat}}
\end{equation}
where $N_{Cat}$ is the number of unique categories of country $c$, $p_{kc}=\frac{V_{kc}}{\sum_{k=1}^{N_{Cat}}V_{kc}}$ and $V_{kc}$ is the number of events created by the country $c$ in the category $k$. A low $D_{Cat}$ score indicates that the country's events are focused on only a few categories. A high value of $D_{Cat}$, on the other hand, indicates that a country's number of events is evenly spread throughout all categories. 

\subsection{Measuring Category Prevalence}
We create a matrix that shows how common each category is in each country. Each element in this matrix, $M_{kc}$, represents the normalized rank of category $k$ in country $c$. First, we select a country with $k\geq 10$, resulting in 63 countries to be evaluated further. We sort each country's top-10 categories and then use the union to combine the results. This results in 28 top categories, which are found in 63 countries. Each category is sorted in ascending order from the highest to the lowest for each country, based on the total event count (TEC$_{k}$). We make the implicit assumption that a larger TEC$_{k}$ value is preferred. As a result, the category $k_{i}$ is ranked 1 for each country, with $TEC_{ki}>TEC_{ki'}\forall i',i,i'\in\{1,2,...,n\}, i\neq i'$ is ranked 1. Any missing category is given a rank of 28, indicating that it is not considered important. Formally, a pseudo-code for creating a category rank matrix is summarized in Algorithm~\ref{alg1}.

\begin{algorithm}
\caption{Create category rank matrix}
\label{alg1}
\begin{algorithmic}
\Require country $c$, category $k$, total event count of each category $TEC_{k}$.
\State {Previous top-10, $k_{A}=()$}
\For {$c=1,2,...,90$}
\If{$k\geq 10$}
\State Sort $TEC_{k}$ in ascending order
\State Current top-10, $k_{B}$ $\gets$ get top-10
\State Top-10 $k$ $\gets$ $k_{A}$ $\cup$ $k_{B}$
\State $k_{A}$ $\gets$ $k_{B}$
\EndIf
\EndFor
\State $M_{TEC}$ $\gets$ Top-10 $k$ $\Join$ $TEC_{k}$
\State $M_{kc}$ $\gets$ Assign rank to each column of $M_{TEC}$
\end{algorithmic}
\end{algorithm}

\section{Data}
\label{sec:data}
\subsection{Data Description}
\begin{figure*}[ht!]
    \centering
    \includegraphics[width=1\textwidth]{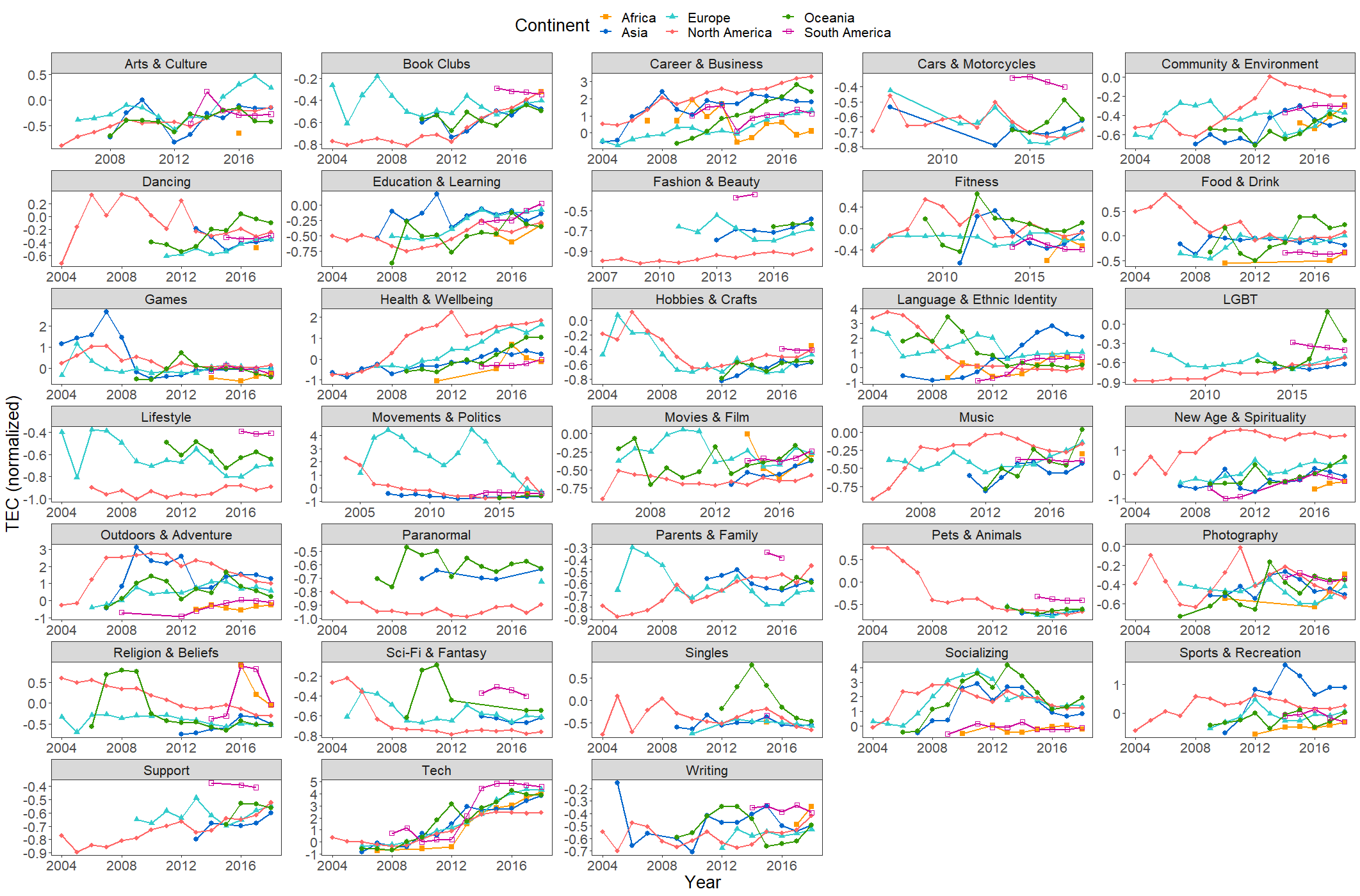}
    \caption{Cumulative number of events (TEC) for each category in six major continents across the timeline. Note that TEC is normalized with z-scores.}
    \label{continentplot}
\end{figure*}

The data used for this study is comprised of Meetup events from 146 countries gathered from February 13, 2003, to May 5, 2019, which was collected at the City University of New York using the Meetup API by the author. The data includes the date, category, geographic location, and short text descriptions posted by event organizers. A total of 2,635,724 event samples were recorded. To narrow this study's scope, we look at the top-90 countries in terms of total event count (TEC), in which a country having fewer than 18 events during the designated period is omitted. Finally, a total of 2,635,508 event samples were included for further analysis. Figure~\ref{continentplot} shows an example of the spatiotemporal evolution of Meetup events across time. A total of 33 different categories were covered in only four countries, while other countries had fewer (see Table~\ref{datasum}). A list of 33 categories, including the corresponding mid- and top-level categories, is presented in Table~\ref{meetupcat}. Table~\ref{datasum} summarizes the dataset used in the analysis.

\subsection{Normality of the Data Distribution}
\begin{figure*}[ht!]
    \centering
    \includegraphics[width=0.7\textwidth]{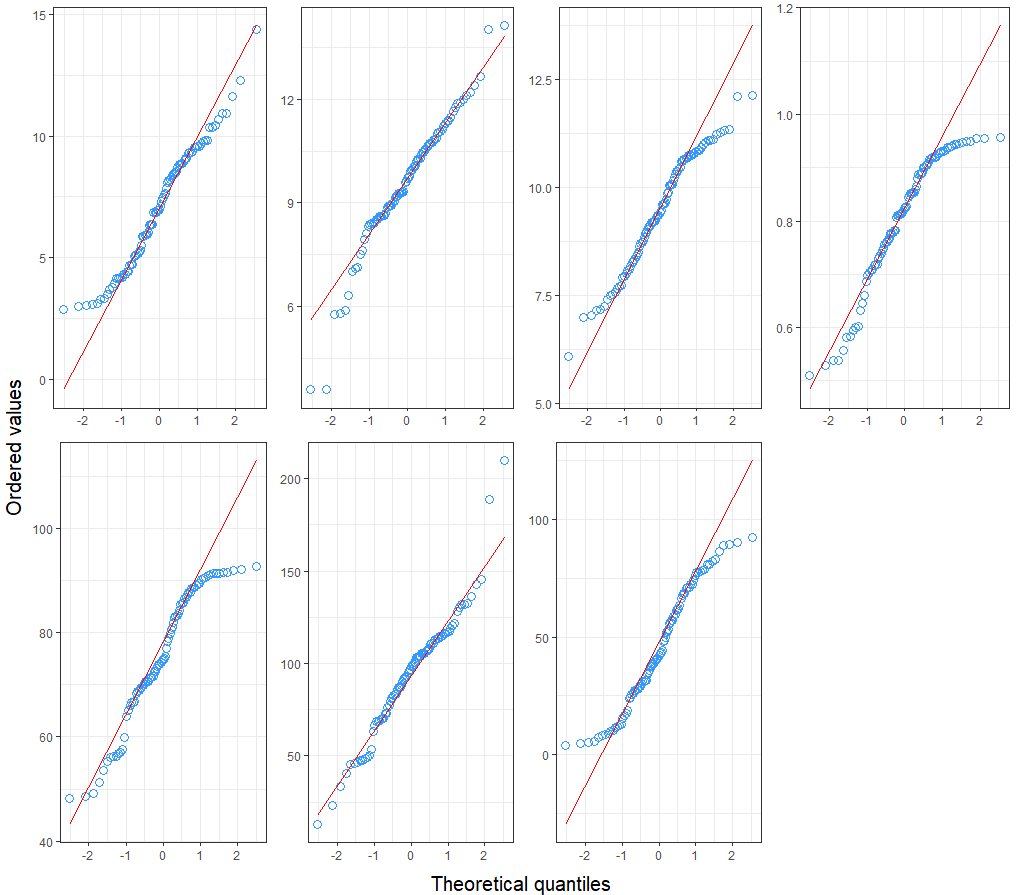}
    \caption{Q-Q plots of logarithmic TEC, logarithmic population, logarithmic GDP, HDI, SPI, MSubs, and IntUs.}
    \label{qqplot}
\end{figure*}

We examine the normality of logarithmic total event count (TEC), logarithmic population, logarithmic GDP per capita (GDP), human development index (HDI), social progress index (SPI), mobile cellular subscriptions per 100 people (MSubs), and individuals using the Internet as a \% of the population (IntUs) of the top-90 countries. The tests indicate that TEC and GDP follow a normal distribution, while population, HDI, SPI, MSubs, and IntUs do not follow a normal distribution. The $p$-values of the Jarque–Bera test for TEC, population, GDP, HDI, SPI, MSubs, and IntUs are 0.417, 0.0016, 0.204, 0.014, 0.091, 0.004, and 0.081, respectively. Hence, we fail to reject the null hypothesis that the data is normally distributed for TEC and GDP but not for the population, HDI, SPI, MSubs, and IntUs. The linear Q-Q plot (see Figure~\ref{qqplot}) confirms the normality of TEC and GDP distributions. 

\section{Results}
\label{find}

\subsection{Association Between Event Dynamics and Socioeconomic Performances}
\begin{figure*}[ht!]
    \centering
    \includegraphics[width=0.8\textwidth]{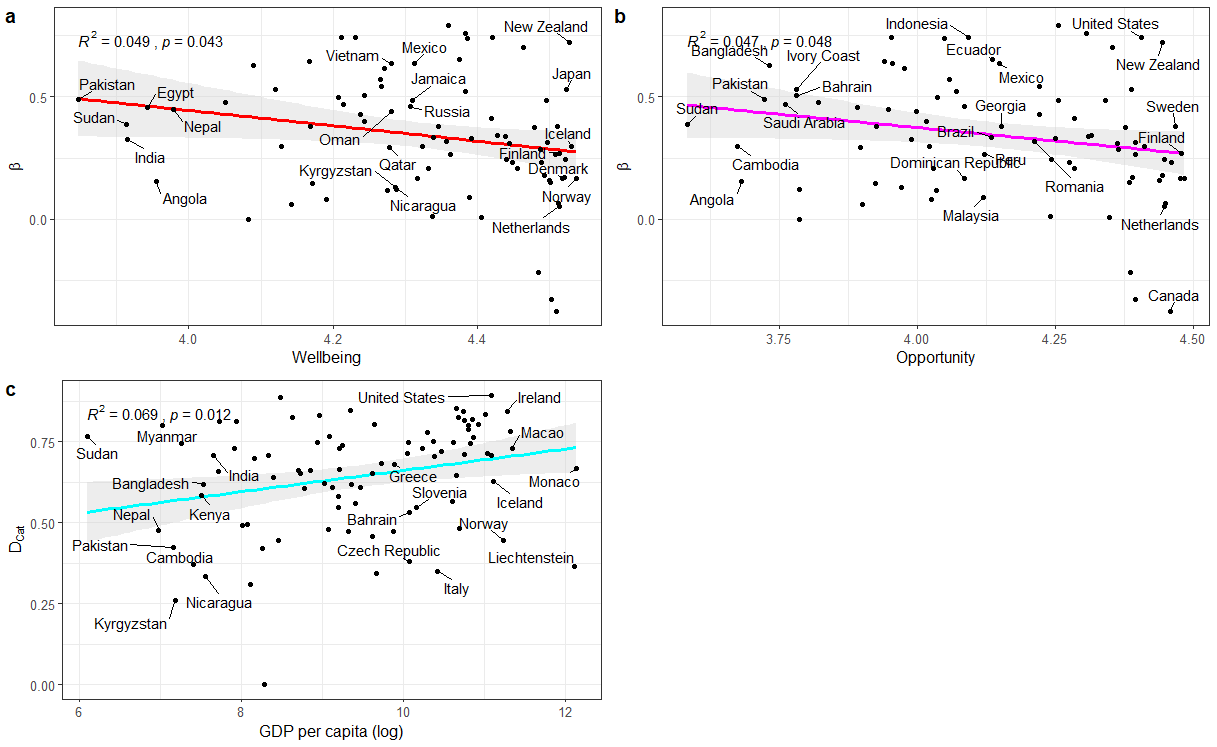}
    \caption{There is a link between event dynamics and socioeconomic indicators. Each panel displays Pearson's $R$ and $p$ values for the correlation test between the relevant pair of variables. a. Association between log wellbeing score and bursty dynamics ($n=83$). b. Association between opportunity and bursty dynamics ($n=83$). c. Association between log GDP per capita and the Meetup categories' diversity index ($n=90$).} 
    \label{soceccorr}
\end{figure*}

Variations in event dynamics patterns are strongly linked to disparities in economic conditions between countries. These correlations provide support for the external validity of the Meetup.com platform. As an illustration, consider that overall, temporal, and category-related traits positively correlate to all socioeconomic indicators except the bursty dynamics. We observe that bursty dynamics negatively correlate with prosperity (as indexed by GDP per capita and human development index (HDI)~\cite{imf2019}), social progress imperative (as scored by basic human needs, foundations of wellbeing, and opportunity~\cite{spi2020}), and digital development index (as measured by the number of internet users per 100 people~\cite{itu2020}) (see Appendix). We highlight three significant relationships that reveal intriguing findings. First, Figure~\ref{soceccorr} reveals a significant association ($R$= -0.22, $p<0.05$) between bursty dynamics ($\beta$) and foundations of wellbeing. Second, bursty dynamics and opportunity score based on the social development index (Figure~\ref{soceccorr}b) have a substantial link ($R$= -0.22, $p<0.05$). Finally, we discover that wealth (as assessed by GDP per capita~\cite{imf2019}) has a positive correlation ($R=0.26$, $p<0.05$) with the diversity index of categories ($D_{Cat}$) (Figure~\ref{soceccorr}c).

\subsection{Cultural Traits versus Event Dynamics}
Features revealed by event dynamics demonstrate a strong correlation to cultural differences according to Hofstede's Cultural Scale~\cite{hofstede2005cultures} and Ingelhart-Welzel World Value Survey~\cite{inglehart2010changing,inglehart2018cultural,inglehart2010wvs}, except that masculinity/femininity (see Appendix). We discuss these correlations of Hofstede's cultural scale as follows. The perceived utility of social networks is directly linked to masculine culture's values of achievement and progress. However, our findings show no substantial relationship between masculine/feminine cultures and event dynamics characteristics. The other five primary cultural attributes and predictors of event dynamics elements are highlighted. First, we notice systematic differences between individualistic and collectivistic communities. Individualistic cultures, which value utility and active engagement in social exchange~\cite{sanchez2009exploring}, show a substantial positive connection to total event count (TEC) and other temporal behavior-related variables (Figure~\ref{cultcorra}a) (see Appendix). 

Second, countries with high degrees of uncertainty avoidance have only a weak negative relationship with the event category variety index ($D_{Cat}$) (Figure~\ref{cultcorra}b). Third, people in low-power distance cultures are more dependent and hence more vulnerable to Meetup.com regardless of power status. Our findings show that bursty dynamics are substantially correlated with high-level power distance cultures (Figure~\ref{cultcorra}b-inset). Fourth, we highlight a collection of correlations involving long-term orientation indices in which specific dynamics properties are significant (see Appendix). We find that the frequency of events and coefficient of variation (CoV) is related in long-term orientation societies (Figure~\ref{cultcorra}c). Finally, we observe that indulgence cultures have significant effects on the average number of events per transaction ($\mu_{Trans}$) and category variety index ($D_{Cat}$) (Figure~\ref{cultcorra}d). 

\begin{figure*}[ht!]
    \centering
    \includegraphics[width=0.8\textwidth]{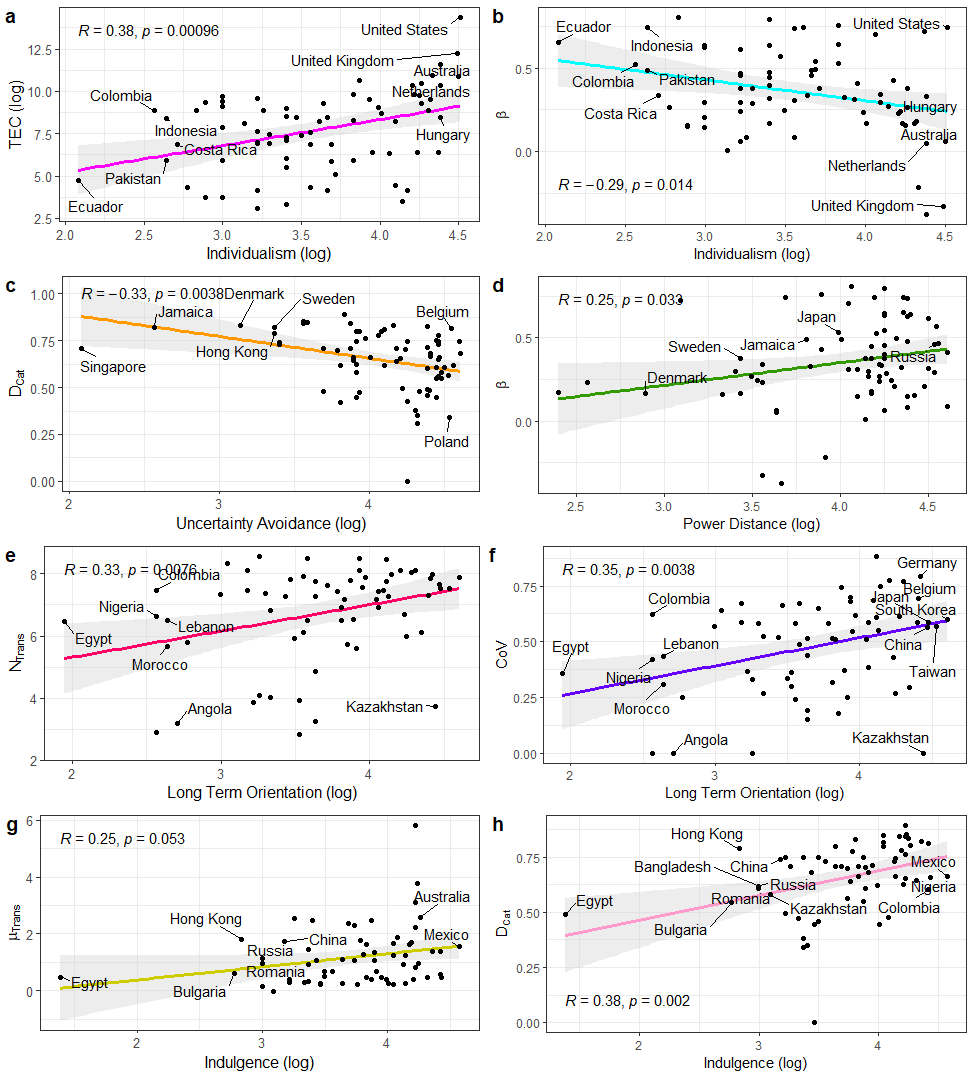}
    \caption{Hoefstede's cultural scales and event dynamics aspects are linked. Each panel displays Pearson's $R$ and $p$ values for the correlation test between the relevant pair of variables. The relationship between log individualism and log total event count (TEC) ($n=74$) (a), or bursty dynamics ($\beta$) ($n=74$) (b). The relationship between log uncertainty avoidance and the category diversity index ($D_{cat}$) ($n=74$) (c) and the relationship between log power distance and bursty dynamics ($n=74$) (d). The relationship between log long term orientation and  log number of transactions ($N_{Trans}$) ($n=66$) (e), or coefficient of variance (CoV) ($n=66$) (f). Relationship between log indulgence and average transaction number of events ($\mu_{Trans}$) ($n=63$) (g), or category diversity index ($D_{cat}$) ($n=63$) (h).} %\mc{=Can we replot this?=}
    \label{cultcorra}
\end{figure*}
 
\subsection{Hierarchical Structure of the Category Prevalence}
\begin{figure*}[ht!]
    \centering
    % \hspace*{-5mm}
    \includegraphics[width=0.8\textwidth]{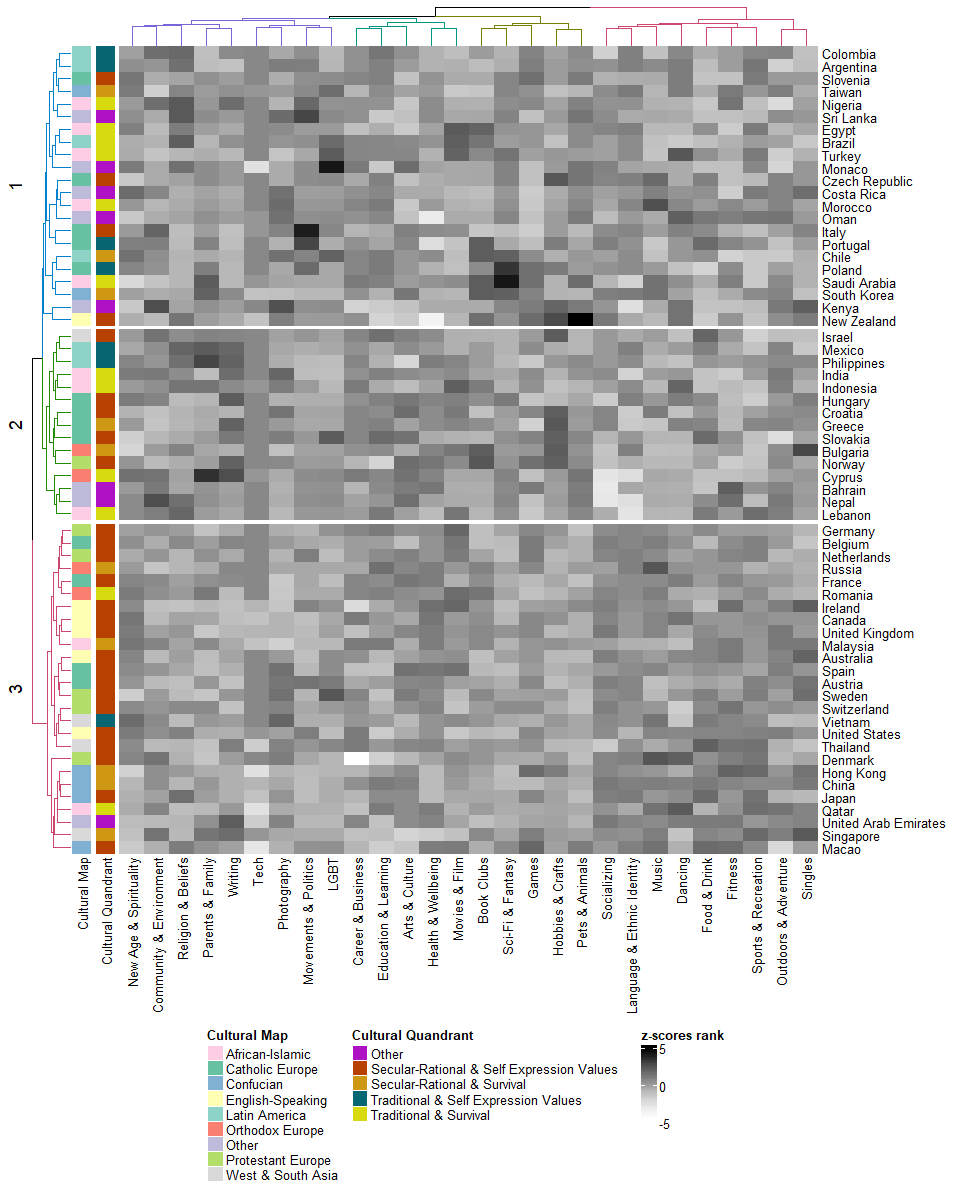}
    \caption{Hierarchical cluster of countries based on category prevalence rank. It is worth noting that the rank is determined using $z$-scores. 63 countries were chosen, each with at least ten categories. Three country-level clusters are represented by the three colors of dendrogram branches. The Inglehart-Welzel World Value Survey 2005-2020 was used to label countries. Significant contrasts can be seen in the differences between clusters. Traditional cultures, for example, dominate Cluster 1. Cluster 3, on the other hand, contains a significant proportion of secular-rational cultures.}
    \label{rankhmap}
\end{figure*}

Based on their geographic location, we can locate countries with homogeneous category preference vectors and explore clusters of countries with such event category predominance. We identified 63 countries with at least ten categories, assigned normalized category ranks to each country, and performed hierarchical clustering using Euclidean distance and Ward's clustering~\cite{murtagh2014ward} (see Methods). Three distinct preference clusters of countries were discovered in this investigation. These are illustrated in Figure~\ref{rankhmap}, and are generally consistent in terms of cultural proximity, according to the Inglehart-Welzel World Value Survey 2005-2020~\cite{inglehart2010changing,inglehart2018cultural,inglehart2010wvs}.     

The first cluster includes African-Islamic, Catholic Europe, Latin America, and other countries not represented on the World Cultural Map. Countries in the cluster also stress Traditional and Survival Values, Traditional and Self Expression Values, and Secular-Rational and Survival Values. The second group comprises Catholic Europe and several Orthodox European countries as well as African Islamic countries and Latin American states. This cluster also includes Traditional and Survival Values and Secular-Rational and Self-Expression Values. English-speaking Protestant Europe and Confucianist cultural groupings such as Macao, Japan, China, and Hong Kong make up the third cluster. Secular-Rational and Self-Expression Values, as well as Secular-Rational and Survival cultural groups, are all included.

\begin{figure*}[ht!]
\centering
\includegraphics[width=0.7\textwidth]{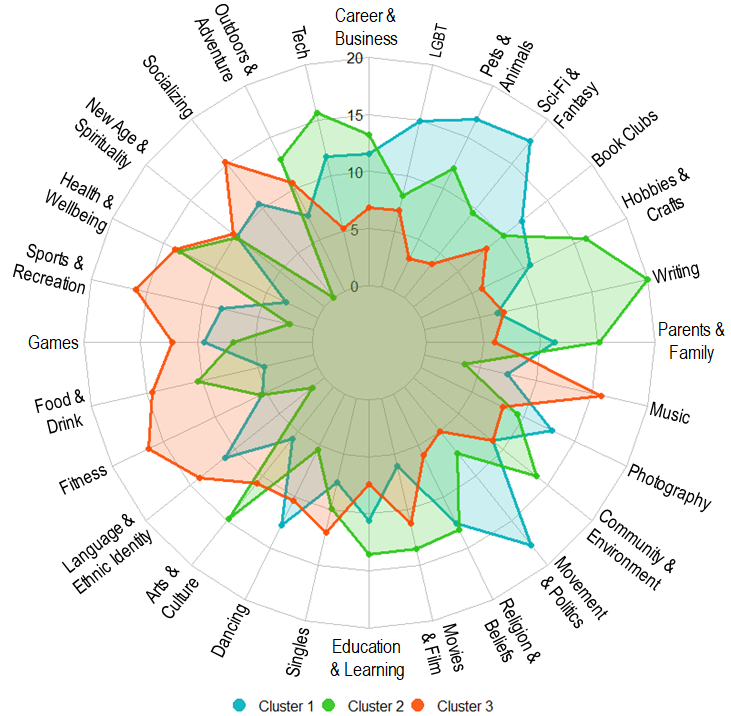}
\caption{Mean rank $z$-scores of the three main clusters as cluster 1, cluster 2, and cluster 3. The mean rank $z$-scores of three clusters is shown as a radar map, revealing an exciting pattern of variations across the clusters throughout the 28 event categories. When compared to countries in other clusters, countries in cluster 3 have a strong preference for leisure-type event categories (such as music, sports and recreation, fitness, and socializing).}
\label{averagerankofeachcluster}
\end{figure*}

\begin{figure*}[ht!]
    \centering
    \includegraphics[width=0.7\textwidth]{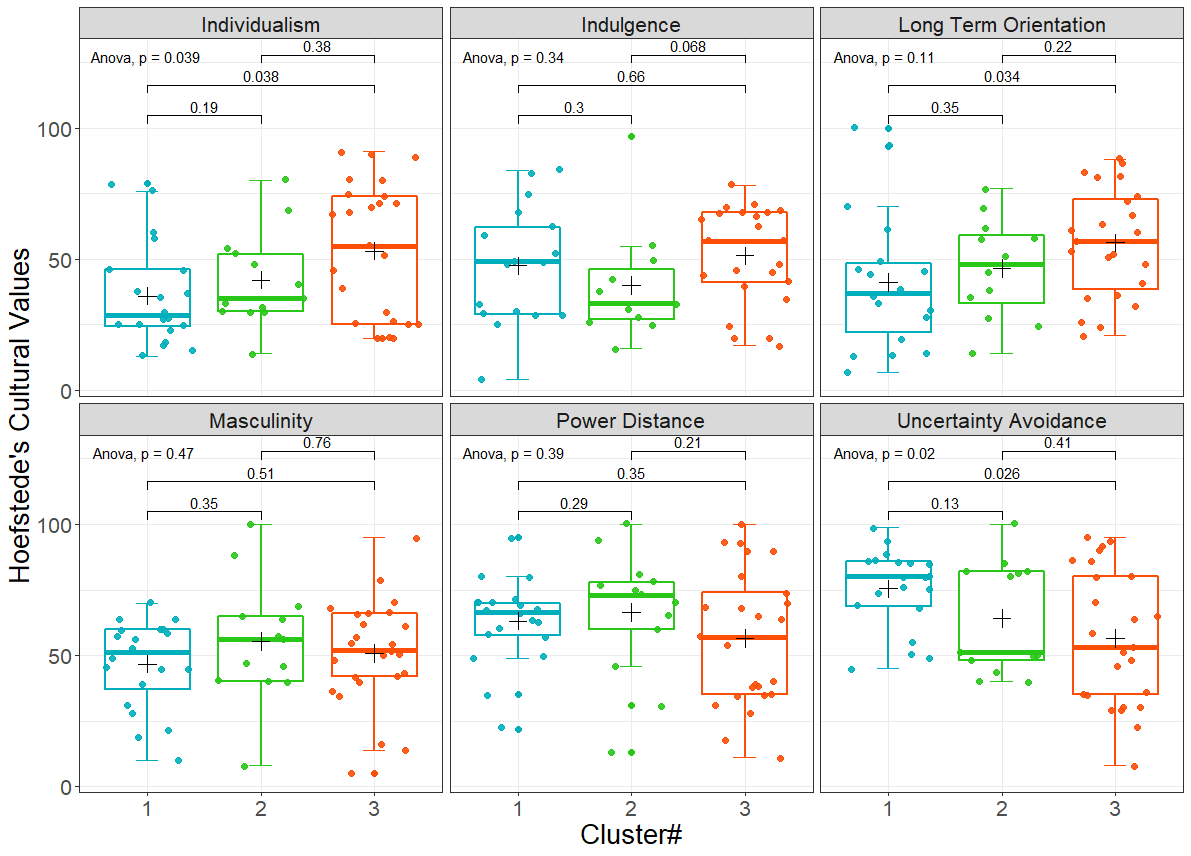}
    \caption{Boxplots (center, median; plus, mean; box, interquartile range (IQR); whiskers, 1.5$\times$IQR; outliers, dots) showing Hofstede's six cultural value dimensions of three major clusters. An omnibus test was calculated using a one-way ANOVA test to compare multiple groups, while pairwise comparisons were calculated using the Wilcoxon test. There are significant differences between clusters. Cluster 1 is mostly made up of civilizations that seek to avoid uncertainty, whereas cluster 3 is made up of countries with high levels of individuality.}
    \label{averageclus}
\end{figure*}

On the other hand, as indicated by cluster 1, traditional cultures have a high interest in movement and politics because they respect national pride and a patriotic viewpoint. In other words, countries in cluster 1 appear to use Meetup.com more frequently to engage in national politics or transformations dialogue. Cluster 2, on the other hand, demonstrates a considerably larger predilection for writing and technology than the other two clusters, despite the lack of cultural proximity. Finally, all clusters appear to have the same modest inclination towards language and identity, new age and spirituality, and singles.

\section{Discussion}
\label{dis}
A study on cultural analytics suggests that data science methods can be used to quantitatively evaluate the patterns, trends, and dynamics of people's intellectual preferences~\cite{michel2011quantitative,kim2014large,manovich2020}. Meetup is a ``digital culture'' as its platform connects physical activities. The main contribution of this study is an in-depth quantitative investigation of Meetup.com's event dynamics patterns and how they connect to cultural attributes. By examining a vast data set from Meetup.com that captures real, cultural events organized by people worldwide, we offer an organic way to identify world cultural features and differences. Our method captures differences in cultural event dynamics and generates a set of features that can be used to quantify cultural regularities statistically across countries throughout the world. Furthermore, it enables us to detect cultural closeness, reflecting groups of countries with a shared interest in specific issues. We tested whether event dynamics patterns help explain socioeconomic indicators across 90 societies on all continents. The implications of our findings are discussed as follows.

First of all, the current study found a negative association between the event consistency and the country's well-being. In other words, the more foundational education, information, and understanding a country's population has and the conditions for living a healthy life, the more probable the country is to have entirely stable event usage. The result confirmed previous findings that the usage of social networking sites might boost well-being by increasing social capital, which in turn enhances social support~\cite{van2016online,trepte2015influence,trepte2016social,utz2017relationship}. While most existing studies focused on pure online social networking sites, this study added a research implication that social support could also happen in an offline setting through face-to-face meeting events. In a similar vein, the opportunity domain of the SPI is negatively associated with its bursty dynamics. Creating a consistent number of Meetup events in a given country is intimately linked to a citizen's right to freedom of speech. This correlation is explained by a society that supports an individual's self-esteem and encourages more self-expression. Furthermore, our correlation showed that a rich country tends to have diverse event categories. The number of events in wealthy countries is not concentrated in a few categories, owing to the enormous number of people who engage in various cultural activities. While Meetup events are considered to represent real-world cultural activities, one could argue that cultural and social involvement has always been a means of encouraging rapid economic development in a given country~\cite{ivaldi2017cultural}.

This current research revealed that individualistic cultures also have a more consistent pattern regarding the number of events ($\beta<0$). This correlation aligns with the existing literature in which individualistic cultures value self-expression (e.g., self-esteem) and independence more than collectivist societies. The Meetup.com platform is based on events organized by individuals who want to share their ideas and communicate with others about similar issues. Even though Meetup comprises social structures and not stranger-to-stranger contacts, motivation is required to create events that will entice people to participate in such activities. In contrast to collectivist societies, the more independent a society is, the more encouraged people are to share their opinions and fearlessly express them. In the countries with a high value of uncertainty avoidance score, we show that such countries have low $D_{Cat}$, indicating the choice of attending the event was limited to a few categories. Because Meetup.com allows for social presence, there is a lesser sense of uncertainty, which makes meeting new people less intimidating. However, as evidenced by the low $D_{Cat}$, we can deduce that uncertainty dampens the impact of shifting to new event categories. For societies with high-level power distance, there is a tendency that the social media platform is difficult to adopt since the number of events has appeared at random ($\beta>1$)~\cite{RN11}. 

% To put it another way, the number of events in high-level power distance cultures has happened at random, owing to the fact that interactive contact on Meetup.com is considered a difficult or unpleasant activity. This is in line with previous research that suggests Meetup.com may be difficult to adopt in countries with high power distance ratings~\cite{RN11}.

Furthermore, our correlation shows that the number of events increases in long-term-oriented societies. We might argue that Meetup.com has been preferred to maintain a long-term connection. Building trust and deep bonds in high short-term oriented societies, on the other hand, is difficult and time consuming~\cite{marcus2009cross}. Meetup.com appears to be an excellent platform for attracting devoted attendees, allowing event organizers to identify and better predict turnout. This study also shows a positive impact of the social media platform on indulging societies. Indulgent societies emphasize enjoying life and having fun, mainly leisure, spending, and consumption. Thus, we can argue that the average number of events per transaction and category variations on Meetup.com are linked to social leisure, offering the members the opportunity to fulfill social needs. This result is in line with previous work mentioning that Meetup.com is a platform that meets Maslow's Hierarchy of Needs for belonging, esteem, and self-actualization~\cite{RN18}.

Next, we discuss the correlations between dynamic pattern features and cultural traits based on the Ingelhart-Welzel World Value Survey. We observe that survival vs. self-expression cultural values possess significant correlations with all pattern dynamics measures, including overall behavior pattern, temporal behavior pattern, and event category-related behavior pattern (see Appendix). In other words, it can be implied that the higher the self-expression values, the larger the production of Meetup.com events. Furthermore, the events were evenly distributed in all available categories, implying that the events were diverse. The low level of trust in a survival-oriented culture may explain why Meetup.com is not widely accepted. An adequate level of trust is required to enable the formation of the social structure~\cite{RN11}. In a society that lacks trust, it is hard to form a corporate structure~\cite{govier1997social}. Consequently, it may be challenging to build a social structure in a survival-oriented society with low-level trust. 

Our findings show that the associations between traditional vs. secular-rational authority values and all dynamic pattern features are substantial, except for bursty dynamics and the category variability index (see Appendix). The significant association between these cultural values and the overall event count, transaction count, or the average number of events per transaction on Meetup.com shows that events are actively generated in more secular cultures. If Meetup.com represents cultural beliefs, it is reasonable to assume that it is rarely used for religious or traditional family purposes. Secular cultures regard religion and traditional family values less. Another important conclusion is that, despite the large number of categories ($N_{Cat}$), events in secular-oriented societies were not evenly distributed throughout the various categories (i.e., $D_{Cat}$ is unimportant).

The clustering pattern suggests that a group of countries' cultural proximity may enable them to acquire common interests in event categories. The differences between the clusters show an interesting pattern of variation. As demonstrated in Figure~\ref{averagerankofeachcluster}. For example, compared to Cluster 1, countries in Cluster 3 have a substantially lower preference for religion and beliefs, movement and politics, and religion and beliefs. Cluster 3 is a secular-rational-oriented culture in which religion, traditional family values, and authority are less important. Cluster 3 countries, on the other hand, are more concerned with personal values such as leisure and the development of interpersonal ties. Music, socializing, sports \& entertainment, and fitness are becoming increasingly popular.

Our clustering pattern also reveals cultural proximity as measured by Hofstede's cultural dimension values. Cluster 1 is mainly made up of countries that mark low in uncertainty and ambiguity values, while Cluster 3 is mostly made up of countries that value individuality, as seen in Figure~\ref{averageclus}. Compared to the other two clusters, Cluster 2 has no specific cultural traits. Uncertainty avoidance cultures tend to construct specific beliefs and norms to prevent uncertain situations, and they do not allow unorthodox behaviors~\cite{smith1992organizational}. \cite{mari2021conspiracy} suggested that the uncertainty avoidance index, in particular, impacted institutional trust. As a result, the prominence of the movement and politics categories in Cluster 1 can be interpreted to be significantly associated with a high level of uncertainty avoidance. Finally, countries with individualistic cultures (e.g., Cluster 3) that value out-group connections above in-group interactions strongly prefer socializing and other leisure-related categories. It could be because people in an individualistic society have a lot of ability to pick and choose their organizations, and there are a lot of them.

\section{Conclusion}
\label{conc}
We investigated the landscape of cultural events held on Meetup.com, an event-based social network that allows organizers to host offline events. Using 17 years of event data, we developed multiple statistical analysis approaches to capture the dynamic event patterns such as the overall behavior pattern, temporal pattern, and category-related pattern. We demonstrated that many event dynamics and category occurrences are strongly correlated with the economic and cultural characteristics of the countries. Furthermore, a clustering technique to uncover the category prevalence was built. The result offered a better understanding of cultural differences, particularly how people from traditional and secular cultures chose their preferences within Meetup's categories.

Subsequently, we also raise several issues and limitations associated with our study. Even though Meetup.com is extensively used worldwide, the event distribution is still heavily skewed towards the United States and specific event categories such as movement and politics in Italy. Furthermore, because social diffusion dynamics drive Meetup's growth directly or indirectly, our data set may indicate greater aggregation and imbalanced distortions. Furthermore, we assume that event categories are disjoint sets, even though they are likely to overlap in the actual world because event organizers may choose them at random when planning a new event. Furthermore, we assume that the total of individual or group cultural values equals country-level values. However, cultural values are unlikely to be universally agreed upon by all the people of a country~\cite{leung2015values,peterson2016culture}. Finally, we may identify how these values are shared across countries by obtaining individual-level data on cultural value orientations and category preferences, allowing for a more nuanced assessment of the extent to which a country implies culture~\cite{gelfand2011differences}. Despite these constraints, we believe that this research can provide a new viewpoint on how event dynamics patterns in social network platform reflect country-level cultural norms. Our method highlights the relationship between culture, nation, and group behaviors, as illustrated on Meetup.com, allowing us to learn more about how cultural events are created, organized, valued, and experienced.

% \begin{acks}
% This research is funded by Institute for Basic Science (IBS) under grant No. IBS-R029-C2-001.
% \end{acks}

\bibliographystyle{ACM-Reference-Format}
\bibliography{samplebase}

\appendix
\section{Data Description}

\begin{table*}[ht!]
\caption{The description of the Meetup dataset in the chosen 90 countries, shown in descending order by total event count (TEC).}
\label{datasum}
\centering
\resizebox{1\textwidth}{!}{ 
\begin{tabular}{|p{2.5cm}|l|l|l|l|l|l|l|l|l|l|l|l|}
\hline
Country & Code & Lat & Lon & \#City & TEC & N$_{Cat}$ &N$_{Trans}$& $\mu_{Population}$ & GDP & HDI   & MSubs&IntUs\\
\hline\hline
United States        & US   & 37.090   & -95.713   & 7,117 & 1,771,002  & 33&5,303  & 309,445      & 65,300    & 0.926 & 92.23&	71.04 \\
United Kingdom       & GB   & 55.378   & -3.436    & 3,090 & 211,273    & 33&4,862  & 63,463       & 42,300    & 0.932 & 116.31&	80.88 \\
Canada               & CA   & 56.130   & -106.347  & 519   & 110,559    & 33&4,951  & 34,300       & 46,200    & 0.929 & 69.64&	78.80  \\
Italy                & IT   & 41.872   & 12.567    & 1,296 & 55,064     & 30&4,698  & 59,439       & 33,200    & 0.892 & 142.90&	48.55\\
Australia            & AU   & -25.274  & 133.775   & 1,277 & 54,466     & 33&4,169  & 22,271       & 55,100    & 0.944 & 100.63&	77.36\\
India                & IN   & 20.594   & 78.963    & 279   & 42,788     & 29&3,316  & 1,238,039    & 2,100     & 0.645 & 47.62&	10.29\\
France               & FR   & 46.228   & 2.214     & 633   & 34,609     & 30&2,958  & 62,932       & 40,500    & 0.901 & 91.88&	67.56\\
Germany              & DE   & 51.166   & 10.452    & 481   & 30,713     & 31&2,932  & 81,539       & 46,400    & 0.947 & 113.39&	78.08\\
Netherlands          & NL   & 52.133   & 5.291     & 330   & 30,646     & 32&3,304  & 16,684       & 52,300    & 0.944 & 112.36&	86.43\\
Japan                & JP   & 36.205   & 138.253   & 113   & 18,116     & 29&1,836  & 128,188      & 40,200    & 0.919 & 99.26&	77.56\\
Switzerland          & CH   & 46.818   & 8.228     & 408   & 17,995     & 31&3,306  & 7,876        & 82,000    & 0.955 & 117.51&	80.98\\
Ireland              & IE   & 53.413   & -8.244    & 139   & 17,283     & 28&3,339  & 4,473        & 78,700    & 0.955 & 103.48&	66.54\\
Singapore            & SG   & 1.352    & 103.820   & 1     & 16,229     & 20&3,139  & 5,047        & 65,200    & 0.938 & 131.72& 71.04 \\
United Arab Emirates & AE   & 23.424   & 53.848    & 13    & 14,486     & 25&2,748  & 7,602        & 43,100    & 0.89  & 145.61 & 68.33   \\
Hong Kong            & HK   & 22.396   & 114.109   & 9     & 14,188     & 29&2,340 & 7,003       & 48,700    & 0.949 &188.81 & 70.68   \\
Spain                & ES   & 40.464   & -3.749    & 181   & 13,798     & 28&2,382 & 45,885       & 29,600    & 0.904 & 105.40  & 63.58  \\
Belgium              & BE   & 50.504   & 4.470     & 214   & 13,753     & 28&2,593 & 10,960       & 46,400    & 0.931 & 103.70 & 72.24   \\
China                & CN   & 35.862   & 104.195   & 60    & 11,799     & 29&2,117 & 1,372,420    & 10,300    & 0.761 & 63.09 & 31.10  \\
South Korea          & KR   & 35.908   & 127.767   & 45    & 11,290     & 24&2,660& 49,770       & 31,800    & 0.916 & 99.08 & 82.19 \\
Brazil               & BR   & -14.235  & -51.925   & 128   & 11,071     & 29&1,708 & 196,276      & 8,720     & 0.765 & 90.48  & 38.94 \\
Sweden               & SE   & 60.128   & 18.644    & 111   & 10,670     & 27&2,673 & 9,436        & 51,600    & 0.945 & 115.33 & 89.24 \\
Thailand             & TH   & 15.870   & 100.993   & 59    & 8,869      & 30&2,470 & 67,238       & 7,810     & 0.777 & 105.34 & 25.77  \\
Israel               & IL          & 31.046   & 34.852    & 98    & 8,588      & 25&2,323 & 7,356        & 43,600    & 0.919 & 121.93 & 57.14 \\
Mexico               & MX          & 23.635   & -102.553  & 96    & 8,051      & 31&1,720 & 114,751      & 9,950     & 0.779 & 68.66  & 33.95  \\
Malaysia             & MY          & 4.210    & 101.976   & 71    & 7,246      & 27&2,083 & 28,270       & 11,400    & 0.81  & 110.21 & 58.58 \\
Denmark              & DK          & 56.264   & 9.502     & 81    & 7,030      & 27&2,711 & 5,568        & 60,200    & 0.94  & 116.24 & 89.02 \\
Taiwan               & TW          & 23.698   & 120.961   & 17    & 6,963      & 26&1,841 & 23,179       & 32,100    & 0.907 &NA  & NA \\
Colombia             & CO          & 4.571    & -74.297   & 43    & 6,951      & 27&1,762 & 45,428       & 6,430     & 0.767 &87.21  & 35.11 \\
Austria              & AT          & 47.516   & 14.550    & 80    & 6,056      & 23&1,732 & 8,477        & 50,100    & 0.922 &136.16 & 72.40 \\
Romania              & RO          & 45.943   & 24.967    & 31    & 5,072      & 21&1,930 & 20,534       & 12,900    & 0.828 &98.46& 38.40 \\
Hungary              & HU          & 47.162   & 19.503    & 39    & 4,834      & 21&1,687 & 9,917        & 16,700    & 0.854 & 111.04& 59.50  \\
Philippines          & PH          & 12.880   & 121.774   & 81    & 4,712      & 26&1,597& 94,902       & 3,490     & 0.718 & 81.82& 24.25 \\
Russia               & RU          & 61.524   & 105.319   & 20    & 4,611      & 21&1,498& 144,206      & 11,600    & 0.824 & 127.92& 43.23 \\
Indonesia            & ID          & -0.789   & 113.921   & 73    & 4,320      & 26&1,442& 243,766      & 4,140     & 0.718 & 83.01& 12.23 \\
Argentina            & AR          & -38.416  & -63.617   & 46    & 3,930      & 25&1,555& 41,169       & 9,910     & 0.845 & 113.46 & 40.72 \\
Turkey               & TR          & 38.964   & 35.243    & 40    & 3,761      & 19&1,299& 73,529       & 9,130     & 0.82  & 82.70 & 37.47\\
Poland               & PL          & 51.919   & 19.145    & 26    & 3,612      & 19&1,441& 38,226       & 15,700    & 0.88  & 114.69 & 55.65 \\
Chile                & CL          & -35.675  & -71.543   & 32    & 3,250      & 19&1,453& 17,183       & 14,900    & 0.851 & 104.02& 49.87  \\
Vietnam              & VN          & 14.058   & 108.277   & 25    & 2,574      & 22&1,292& 88,634       & 2,720     & 0.704 & 91.46& 27.48 \\
Qatar                & QA          & 25.355   & 51.184    & 10    & 2,088      & 22&964& 1,828        & 62,100    & 0.848 &114.43& 57.09 \\
Greece               & GR          & 39.074   & 21.824    & 53    & 1,993      & 21&1,004& 10,893       & 19,600    & 0.888 &105.62& 44.43 \\
Portugal             & PT          & 39.400   & -8.224    & 43    & 1,762      & 22&905& 10,472       & 23,300    & 0.864 &113.17& 51.82 \\
Croatia              & HR          & 45.100   & 15.200    & 22    & 1,663      & 18&995& 4,302        & 14,900    & 0.851 &100.63& 52.15 \\
Bulgaria             & BG          & 42.734   & 25.486    & 11    & 1,475      & 18&802& 7,414        & 9,830     & 0.816 &118.87& 42.01 \\
Nigeria              & NG          & 9.082    & 8.675     & 27    & 1,219      & 13&751& 161,882      & 2,230     & 0.539 & 49.25& 11.91 \\
Saudi Arabia         & SA          & 23.886   & 45.079    & 14    & 1,063      & 21&655& 28,024       & 23,100    & 0.854 &131.90& 43.11 \\
Egypt                & EG          & 26.821   & 30.802    & 32    & 1,030      & 15&635& 84,544       & 3,020     & 0.707 &68.59& 23.82   \\
Slovenia             & SI          & 46.151   & 14.995    & 23    & 1,022      & 11&676& 2,039        & 25,900    & 0.917 &103.09& 61.98  \\
Lebanon              & LB          & 33.855   & 35.862    & 27    & 970        & 15&661& 5,472        & 7,580     & 0.744 &53.57& 40.29 \\
Costa Rica           & CR          & 9.749    & -83.753   & 37    & 950        & 16&550& 4,596        & 12,200    & 0.81  & 86.85& 40.19  \\
Sri Lanka            & LK          & 7.873    & 80.772    & 17    & 938        & 10&672& 20,296       & 3,850     & 0.782 &72.80& 12.91  \\
Slovakia             & SK          & 48.669   & 19.699    & 7     & 593        & 15&450& 5,416        & 19,300    & 0.86  &105.66& 68.63 \\
Norway               & NO          & 60.472   & 8.469     & 12    & 592        & 15&455& 4,934        & 75,400    & 0.957 &107.82& 90.23 \\
New Zealand          & NZ          & -40.901  & 174.886   & 15    & 580        & 15&367& 4,395        & 42,100    & 0.931 &103.88& 76.02 \\
Czech Republic       & CZ          & 49.817   & 15.473    & 13    & 553        & 10&401& 10,481       & 23,500    & 0.9   &120.66& 61.62 \\
Jordan               & JO          & 30.585   & 36.238    & 3     & 414        & 9&330& 7,594        & 4,410     & 0.729 &88.25& 30.36 \\

\hline
\multicolumn{13}{|c|}{\textit{Continued on next page}}\\
\hline
\end{tabular}
}
\end{table*}

\begin{table*}
\resizebox{1\textwidth}{!}{
\begin{tabular}{|p{2.5cm}|l|l|l|l|l|l|l|l|l|l|l|l|}
\hline
\multicolumn{13}{|c|}{\textit{Continued from previous page}}\\
\hline
Country              & Code & Lat & Lon & \#City & TEC & N$_{Cat}$ &N$_{Trans}$& $\mu_{Population}$ & GDP & HDI   &MSubs&IntUs\\
\hline\hline
Bahrain              & BH          & 25.930   & 50.638    & 14    & 391        & 12&238& 1,195        & 23,500    & 0.852 &132.54& 61.22 \\
Morocco              & MA          & 31.792   & -7.093    & 15    & 378        & 14&286& 32,717       & 3,200     & 0.686 & 87.25& 39.28 \\
Pakistan             & PK          & 30.375   & 69.345    & 11    & 364        & 9&273& 181,804      & 1,280     & 0.557 & 48.46& 8.91 \\
Bangladesh           & BD          & 23.685   & 90.356    & 5     & 361        & 5&309& 148,441      & 1,860     & 0.632 &45.22& 5.38 \\
Tunisia              & TN          & 33.887   & 9.537     & 12    & 352        & 5&279& 10,703       & 3,320     & 0.74  &93.12& 31.68 \\
Nepal                & NP          & 28.395   & 84.124    & 19    & 243        & 11&209& 26,695       & 1,070     & 0.602 &45.73& 7.29 \\
Cyprus               & CY          & 35.126   & 33.430    & 5     & 199        & 10&157& 1,106        & 27,900    & 0.887 &111.04& 53.28 \\
Ivory Coast          & CI          & 7.546    & -5.548    & 3     & 191        & 4&173& 20,995       & 2,280     & 0.538 &66.45 & 11.68 \\
Oman                 & OM          & 21.513   & 55.923    & 16    & 174        & 11&142& 3,377        & 15,300    & 0.813 & 116.86& 37.38 \\
Sudan                & SD          & 12.863   & 30.218    & 1     & 170        & 2&141& 35,250       & 442       & 0.51  &46.40 & 15.31 \\
Georgia              & GE          & 42.315   & 43.357    & 2     & 163        & 6&127& 4,105        & 4,700     & 0.812 &85.91& 27.22 \\
Mauritius            & MU          & -20.348  & 57.552    & 11    & 157        & 5&116& 1,244        & 11,100    & 0.806 &96.40& 29.16 \\
Liechtenstein        & LI          & 47.166   & 9.555     & 4     & 112        & 5&68& 36 & 181,000   & 0.919 &96.98& 78.47 \\
Ecuador              & EC          & -1.831   & -78.183   & 7     & 111        & 9&97& 15,151       & 6,180     & 0.759 &77.15& 27.89 \\
Macao                & MO          & 22.211   & 113.553   & 1     & 107        & 12&88& 547          & 84,100    & 0.909 &210.04& 56.25 \\
Nicaragua            & NI          & 12.865   & -85.207   & 3     & 84         & 5&74& 5,869        & 1,910     & 0.66  &73.18& 9.82  \\
Iceland              & IS          & 64.963   & -19.021   & 4     & 83         & 8&56& 317          & 66,900    & 0.949 &107.90& 92.41\\
Peru                 & PE          & -9.190   & -75.015   & 8     & 75         & 9&47& 29,322       & 6,980     & 0.777 &76.26 & 31.61  \\
Algeria              & DZ          & 28.034   & 1.660     & 11    & 74         & 1&60& 36,685       & 3,970     & 0.748 &81.26& 16.40 \\
Monaco               & MC          & 43.750   & 7.413     & 1     & 66         & 10&65& 36           & 186,000   & NA     &70.50& 73.90  \\
Kenya                & KE          & -0.024   & 37.906    & 5     & 65         & 11&55& 42,769       & 1,820     & 0.601 &50.10& 8.40 \\
Jamaica              & JM          & 18.110   & -77.298   & 9     & 64         & 7&50& 2,821        & 5,580     & 0.734 &95.12& 27.07 \\
South Africa         & ZA          & -30.559  & 22.938    & 7     & 64         & 9&50& 51,902       & 6,000     & 0.709 &106.63& 25.53 \\
Cuba                 & CU          & 21.522   & -77.781   & 7     & 51         & 8&44& 11,273       & 8,820     & 0.783 &12.84& 18.85  \\
Angola               & AO          & -11.203  & 17.874    & 1     & 43         & 3&24& 24,036       & 2,790     & 0.581 &33.59& 4.73 \\
Kazakhstan           & KZ          & 48.020   & 66.924    & 1     & 41         & 5&42& 16,534       & 9,810     & 0.825 &107.30& 35.96 \\
Finland              & FI          & 61.924   & 25.748    & 8     & 33         & 7&26& 5,378        & 48,800    & 0.938 &130.13& 83.19 \\
Cambodia             & KH          & 12.566   & 104.991   & 2     & 28         & 2&15& 14,488       & 1,640     & 0.594 &70.11& 8.11 \\
Dominican Republic   & DO          & 18.736   & -70.163   & 4     & 27         & 6&18& 9,749        & 8,280     & 0.756 &69.47& 31.04 \\
Kyrgyzstan           & KG          & 41.204   & 74.766    & 1     & 23         & 2&23& 5,543        & 1,310     & 0.697 &79.66& 17.66 \\
Tanzania             & TZ          & -6.369   & 34.889    & 5     & 22         & 5&17& 45,522       & 1,120     & 0.529 &40.28& 3.88  \\
Bahamas              & BS          & 25.034   & -77.396   & 5     & 21         & 7&21         & 354          & 34,900    & 0.814 & 83.68& 48.01 \\
Myanmar              & MM          & 21.914   & 95.956    & 2     & 20         & 3&10         & 50,920       & 1,410     & 0.583 &23.14& 5.18 \\
Belize               & BZ          & 17.190   & -88.498   & 5     & 18         & 6&15         & 326          & 4,820     & 0.716 & 47.66& 29.02 \\
\hline
\end{tabular}
}
\end{table*}

\begin{table*}[ht!]
\caption{Category names available in Meetup. The 33 categories are grouped into 11 mid-level categories and into 7 top-level categories.}
    \label{meetupcat}
    \centering
    \begin{tabular}{|l|l|l|l|}
    \hline
    No&Category&Mid-level category&Top-level category\\
    \hline\hline
   
    1  & Movies \& Film              & \multirow{2}{*}{Cinema and Video Art} & \multirow{5}{*}{Art}                     \\
    2  & Sci-Fi \& Fantasy           &                                       &                                          \\\cline{1-3}
    3  & Dancing                     & Dancing                               &                                          \\\cline{1-3}
    4  & Arts \& Culture             & General Art                           &                                          \\\cline{1-3}
    5 & Music                       & Music and Audio Art                   &                                          \\\hline
    6  & Career \& Business          & Business and Finance                  & \multirow{4}{*}{Training}                \\\cline{1-3}
    7  & Education \& Learning       & Education and Science                 &                                          \\\cline{1-3}
    8 & Tech                        & Technology                            &                                          \\\cline{1-3}
    9 & Writing                     & \multirow{20}{*}{Languages}           &                                          \\\cline{1-2}\cline{4-4}
    10  & Book Clubs                  &                                       & \multirow{15}{*}{Everyday Culture}       \\
    11  & Cars \& Motorcycles         &                                       &                                          \\
    12  & Community \& Environment    &                                       &                                          \\
    13 & Food \& Drink               &                                       &                                          \\
    14 & Games                       &                                       &                                          \\
    15 & Health \& Wellbeing         &                                       &                                          \\
    16 & Hobbies \& Crafts           &                                       &                                          \\
    17 & Lifestyle                   &                                       &                                          \\
    18 & Outdoors \& Adventure       &                                       &                                          \\
    19 & Paranormal                  &                                       &                                          \\
    20 & Parents \& Family           &                                       &                                          \\
    21 & Pets \& Animals             &                                       &                                          \\
    22 & Photography                 &                                       &                                          \\
    23 & Socializing                 &                                       &                                          \\
    24 & Support                     &                                       &                                          \\\cline{1-2}\cline{4-4}
    25 & Language \& Ethnic Identity &                                       & \multirow{4}{*}{Religion and Traditions} \\
    26 & LGBT                        &                                       &                                          \\
    27 & New Age \& Spirituality     &                                       &                                          \\
    28 & Religion \& Beliefs         &                                       &                                          \\\hline
    29 & Singles                     & Other                                 & Other                                    \\\hline
    30 & Movements \& Politics       & Politics                              & Politics                                 \\\hline
    31 & Fashion \& Beauty           & \multirow{3}{*}{Sports}               & \multirow{3}{*}{Sports}                  \\
    32 & Fitness                     &                                       &                                          \\
    33 & Sports \& Recreation        &                                       &                                          \\\hline
    
    \hline
    \end{tabular}
\end{table*}

\section{Overall Correlations}
\begin{figure*}[ht!]
    \centering
    \includegraphics[width=0.82\textwidth]{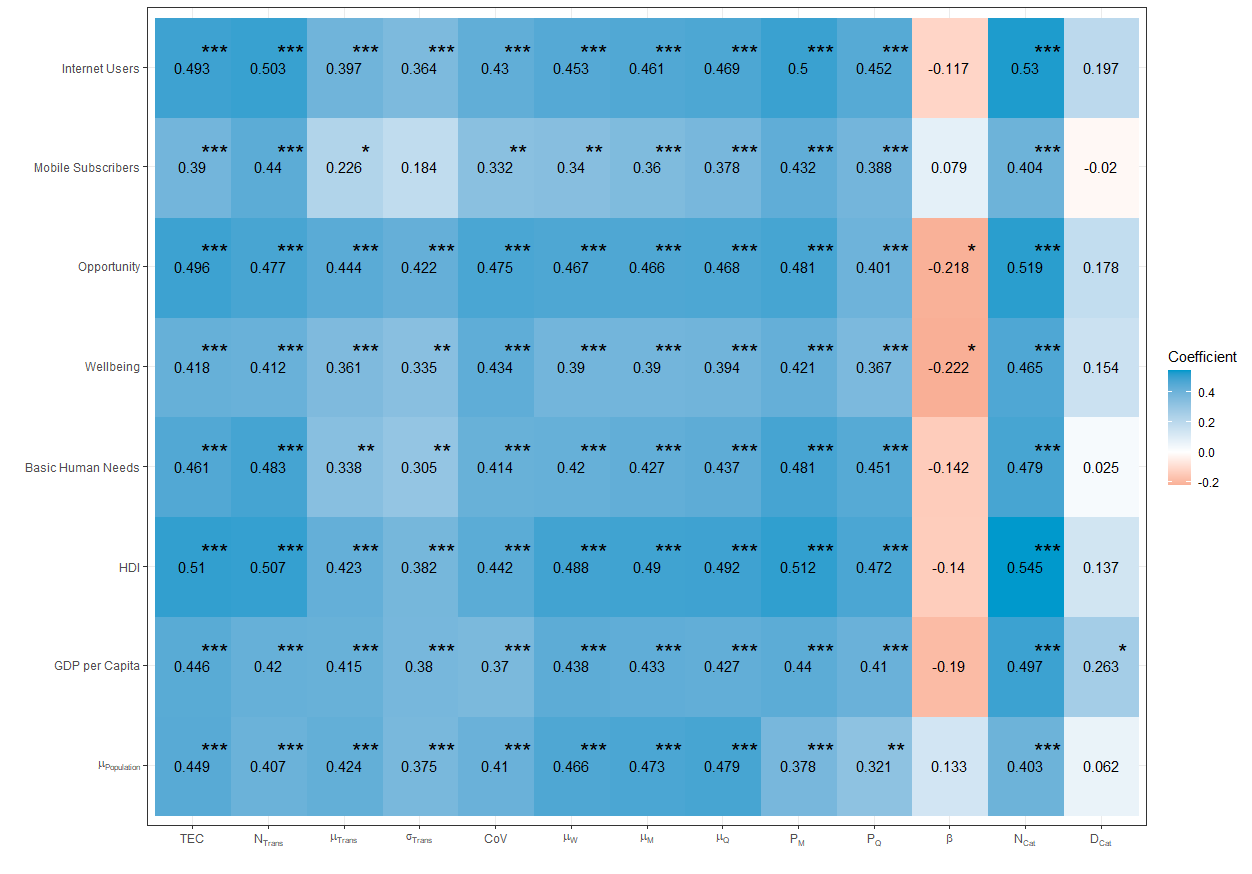}
    \caption{Pearson correlation between event dynamics features and socioeconomic and technology indicators. We use the following notation to denote statistical significance: $*p<0.05$, $**p<0.01$, $***p<0.001$.}
    \label{overallsoceccorr}
\end{figure*}

\begin{figure*}[ht!]
    \centering
    \includegraphics[width=0.82\textwidth]{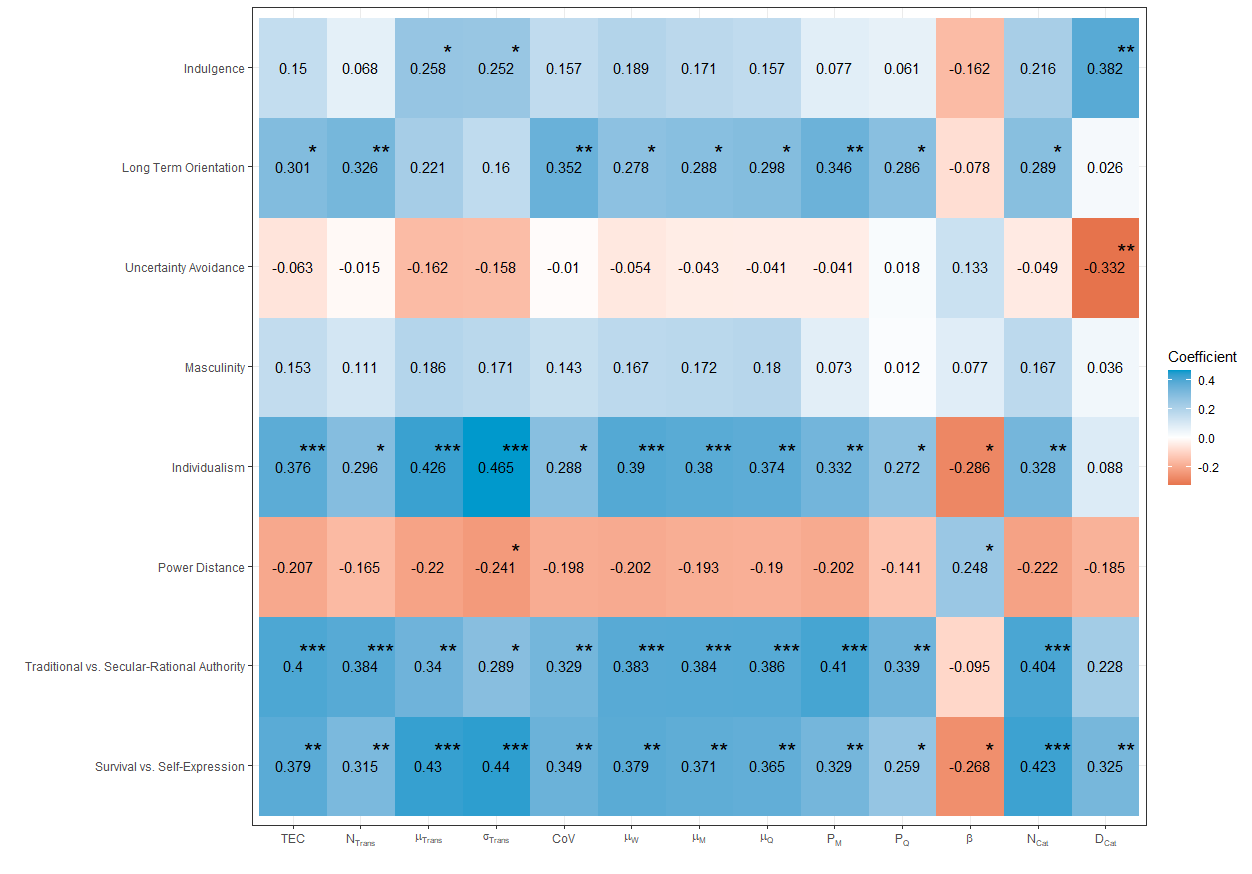}
    \caption{Pearson correlation between event dynamics features and cultural traits based on Hoefstede's Cultural Values and Inglegart-Welzel's World Value Survey. We use the following notation to denote statistical significance: $*p<0.05$, $**p<0.01$, $***p<0.001$.}
    \label{overallcultcorr}
\end{figure*}

In this section, we provide complete correlation results as shown in Figure~\ref{overallsoceccorr} and \ref{overallcultcorr}. 

\end{document}